\documentclass[%
 aip,
 amsmath,amssymb,
 reprint,%
]{revtex4-1}

\usepackage{dcolumn}
\usepackage{bm}
\usepackage{graphicx}
\usepackage{wrapfig}
\usepackage{amsmath}
\usepackage{xspace}
\usepackage{capt-of}
\usepackage{float} 
\usepackage{epsfig}
\usepackage{rotating}
\usepackage{fancyhdr}
\usepackage[titletoc]{appendix}
\usepackage[english]{babel}
\usepackage{subcaption}
\usepackage{hyperref}
\usepackage[dvipsnames]{xcolor}

\usepackage[utf8]{inputenc}
\usepackage[T1]{fontenc}
\usepackage{mathptmx}
\usepackage{xcolor}

\begin{document}

\preprint{AIP}

\title{Determination of Thermal Conductivity of phase pure 10H-SiC Thin Films by non-destructive Raman Thermometry}

\author {Madhusmita Sahoo}
\email{msahoo@igcar.gov.in, krushna.research@gmail.com}
\affiliation{Surface and Sensors Studies Division, Materials Science Group, A CI of HBNI, IGCAR, Kalpakkam 603102, India 
}%
\author{Kalyan Ghosh}%
\affiliation {School of Physical Sciences, National Institute of Science Education and Research (NISER) Bhubaneswar, An OCC of Homi Bhabha National Institute, Jatni-752050, Odisha, India 
}%
\author{Swayamprakash Sahoo}%
\affiliation {School of Physical Sciences, National Institute of Science Education and Research (NISER) Bhubaneswar, An OCC of Homi Bhabha National Institute, Jatni-752050, Odisha, India 
}%

\author{Pratap K. Sahoo}%
\email{pratap.sahoo@niser.ac.in}
\affiliation {School of Physical Sciences, National Institute of Science Education and Research (NISER) Bhubaneswar, An OCC of Homi Bhabha National Institute, Jatni-752050, Odisha, India 
}%
\affiliation {Center for Interdisciplinary Sciences (CIS), NISER Bhubaneswar, HBNI, Jatni-752050, Odisha, India
}%
\author {Tom Mathews}
\affiliation{Surface and Sensors Studies Division, Materials Science Group, A CI of HBNI, IGCAR, Kalpakkam 603102, India 
}%
\author {Sandip Dhara}
\affiliation{Surface and Sensors Studies Division, Materials Science Group, A CI of HBNI, IGCAR, Kalpakkam 603102, India 
}%
\date{\today}
\begin{abstract}
10 H SiC thin films are potential candidates for devices that can be used in high temperature and high radiation environment. Measurement of thermal conductivity of thin films by a non-invasive method is very useful for such device fabrication. Micro-Raman method serves as an important tool in this aspect and is known as Raman Thermometry. It utilises a steady-state heat transfer model in a semi-infinite half space and provides for an effective technique to measure thermal conductivity of films as a function of film thickness and laser spot size. This method has two limiting conditions i.e. thick film limit and thin film limit. The limiting conditions of this model was explored by simulating the model for different film thicknesses at constant laser spot size. 10H SiC films of three different thicknesses i.e. 104, 135 and 156 nm were chosen to validate the thin film limiting condition. It was found that the ideal thickness at which this method can be utilised for calculating thermal conductivity is 156 nm. Thermal conductivity of 156 nm film  is found to be 102.385 $(Wm^{-1}K^{-1})$.

\end{abstract}
\maketitle
\section{INTRODUCTION}
Energy harvesting utilising ionising and non ionising radiation sources have generated a lot of interest in the scientific community in recent years. Amongst various methods and materials, it is particularly challenging to find a suitable material for utilising ionising radiation. SiC has been considered as a suitable material in this regard due to its ability to perform in high temperature and high radiation conditions where conventional semiconductor devices cannot perform adequately.\cite{nava2008silicon} It is important that the thermal conductivity is measured by a non destructive method after the whole device is assembled. It becomes all the more important for an radioactive environment, where one would need necessary provisions and methodology to measure thermal conductivity intermittently to know the health of the device. While many studies have been done for more common polytype like 3C, 4H, and 6H, higher hexagonal polytypes of SiC remain unexplored.\cite{Cheng2022, slack1964thermal, PROTIK201731}  Hence, it is  desirable that a phase pure SiC is synthesized so that defect-phonon interaction can be minimised leading to higher thermal conductivity.\cite{Cheng2022} In this study, we have focused on determining the thermal conductivity of phase pure 10H SiC. Several methods have been reported so far to determine thermal conductivity of dielectric materials. These methods are steady state method,\cite{zhang1995thermal} 3 $\omega $ method,\cite{cahill1994thermal} photo acoustic method,\cite{swimm1983photoacoustic} thermal microscopy method \cite{callard1999thermal} and thermo-reflectance method. \cite{burzo2002influence, komarov2003transient}  These methods are invasive in nature wherein the original sample is either damaged or extensive sample preparation and data analysis is required. Hence, Perichon et al. demonstrated a non invasive and non destructive micro-Raman method to determine thermal conductivity of thick films. \cite{perichon2000} However, necessary condition for this method is the sample thickness must be one order higher than the laser diameter. This implied that only the films having thickness in the order of microns were suitable for the method proposed by Perichon et al.\cite{perichon2000} Huang et al. modified the methodology to evaluate thermal conductivity of thin films having thickness in submicrometer to nanometer range.\cite{Huang2009} Prior to this study, only few researchers  have used the method proposed by Huang et al. for determining thermal conductivity of Si based devices, 2D graphene and biological samples.\cite{gan2015raman,freedman2017substrate,zhang2014raman}  In this report, we have used the method  proposed by Huang et al. for determining thermal conductivity of  10H SiC thin films in the range of 104-156 nm, which is being reported for the first time to the best of our knowledge.
\vspace{-0.34in}
\section{EXPERIMENTAL DETAILS}

SiC thin films were deposited on cleaned Si substrates using RF magnetron sputtering. The deposition parameters such as RF power, deposition time, gas flow, target to substrate distance are as follows. RF power = 90 Watt, reflectant power = 0 Watt, deposition time = 30 mins, gas flow(Ar) =15 sccm and target to substrate distance (3.5inch). Thickness of films were varied by changing only chamber pressure during deposition. Three different thickness of 104 nm, 153 nm and 156 nm were obtained at a chamber pressure of 1*10$^{-2}$, 2*10$^{-2}$ and 5*10$^{-2}$ mbar, respectively. The thickness of the films were measured by cross sectional field emission scanning electron microscopy (FESEM) (SIGMA model, Carl Zeiss). Crystallographic phase of the films were determined from GIXRD by using Rigaku Smart Lab X-ray diffractometer with monochromatic Cu K-$\alpha$ radiation ($\lambda$ = 1.5418 \AA).  Raman spectra of the samples were acquired using a 532 nm laser as the excitation source (Jobin-Yvon LabRam HR Evolution, Horiba). The spectra were collected using a spectrometer coupled with a CCD based detector in back-scattered geometry with 50x lens. Temperature dependent Raman spectra were recorded using a high temperature Linkam stage.\\
\vspace{-0.4in}

\section{Crystallographic and Morphological Studies}
The crystallographic phase of the RF magnetron sputtered films were determined by matching peaks of  X-ray diffraction pattern with peaks from ICDD card no 89-2214. A representational grazing incidence X-ray diffraction pattern of 156 nm thick 10H SiC  is shown in figure 1. The peak at 44.5$^{\circ}$ and 28.3$^{\circ}$ corresponds to (018) and (008) planes of 10 H SiC. 
\begin{figure}[htbp]
\begin{center}
 \includegraphics[width=0.45\textwidth]{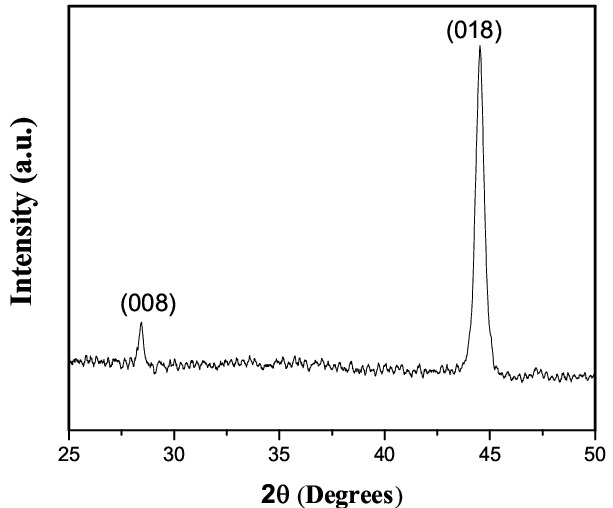}
 \end{center}
\caption{Grazing incidence X-ray diffraction spectra of 156 nm thick 10H SiC film }
\end{figure}
 Surface morphology of 156 nm 10H SiC film is shown in fig 2, which indicated smooth and continuous film formation. All other films were also found to possess similar morphological features. Cross sectional FESEM is shown in inset of fig 2. It reveals uniform thickness across the substrate. The thicknesses measured from cross-sectional FESEM is used in calculation of thermal conductivity of the film (k$_f$), which discussed in detail in Section IV.
\begin{figure}[ht]
\begin{center}
 \includegraphics[width=0.45\textwidth]{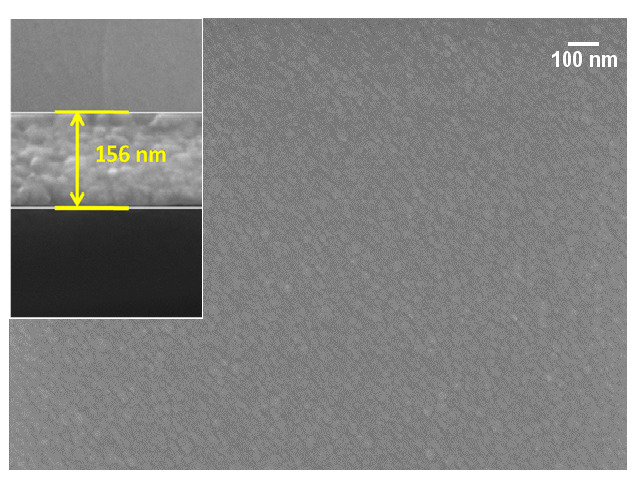}
 \end{center}
\caption{Surface morphology of SiC samples using FESEM (Inset) Cross-sectional thickness of a deposited SiC sample. }
\end{figure}

\section{Raman Thermometry}
\subsection{The model}
It has been a general practice for researchers to combine temperature dependent Raman peak shift within the ambit of thermal scanning probe microscopy and calculate thermal conductivity. However, heating source in both the methods are different. Therefore in this work we have followed the methodology developed by Huang et al. that involves a separate mathematical model to account for both the local heat induced due to a Gaussian laser beam and the thickness of the sample.\cite{Huang2009} 

In this method, a Gaussian laser beam is considered to be incident on a relatively thick substrate. It is assumed that heat transfer is complete and heat loss is minimum. This reduces the problem to a steady state heat transfer problem from a circular region in a semi-infinite space.

The temperature across the film sample then satisfies the Laplacian
\begin{equation}
    \triangledown ^{2}t(r,z) = 0
\end{equation}

Applying necessary boundary conditions Huang et al. deduced the formula for thermal conductivity of the thick film to be 
\begin{equation}
    k = \frac{\left [ I_0\left ( 1 \right ) + I_1\left ( 1 \right ) \right ]\cdot P}{\sqrt{2\pi}er_0\triangle T}
\end{equation}
where $I_0$ and $I_1$ are modified Bessel's equations of zeroth and first order respectively, which are derived by Dryden  P represents power of laser, $r_0$ is radius of laser beam and $\triangle T$ is  induced temperature rise in sample due to laser irradiation. \cite{Dryden} 
The presence of laser beam as a heating source is quintessential to the model proposed. Hence, it becomes necessary to know exact spot size of the laser beam on the sample. The spot size for a Gaussian laser beam is calculated by equation 3, which is given below.
\begin{equation}
    r_o = \frac{\lambda}{\pi \cdot N.A.}
\end{equation}
where $\lambda$ represents the wavelength of the laser beam and N.A. is the numerical aperture of the lens, $r_o$ is required radius of the laser spot.

However, above basic equation focuses on bulk films with thickness at least one order of magnitude higher than the laser beam diameter. Extending this to a thin film model, thermal conductivity of the film in eq. 2 became apparent thermal conductivity ($k_{app}$) of the entire sample as given below.
\begin{equation}
    k_{app} = \frac{\left [ I_0\left ( 1 \right ) + I_1\left ( 1 \right ) \right ]\cdot P}{\sqrt{2\pi}er_0\triangle T}
\end{equation}

Heat flow across the sample must still satisfy equation 1. Hence, a Laplacian was applied to substrate and film separately. Subsequently, necessary boundary conditions were applied to find a relation between thermal conductivity of the substrate($k_s$), thin film($k_f$) and the entire sample($k_{app}$) giving rise to the following equation.

\begin{equation}
    \frac{k_s}{k_{app}} = 1 + \sqrt{\frac{2}{\pi}}\frac{e-e^{-1}}{\left [ I_0\left ( 1 \right ) + I_1\left ( 1 \right ) \right ]}
    \frac{\delta}{r_0}\frac{k_s}{k_f}\left [ 1-\left ( \frac{k_f}{k_s} \right )^2 \right ]
\end{equation}
where $\delta$ is the thickness of the film. 

Equation 5 shows a dependency on $\delta/r_o$ ratio of the sample. Huang et al. have further simplified equation 4 by assuming $k_f$ is less than $k_s$. This holds true for SiO$_2$, which was the material of interest to them. However, there is no literature value available for bulk 10H SiC. Hence, a simulation was carried out using C programme to explore variation of k$_f$ and k$_{app}$ with a change in $\delta/r_o$ ratio to circumvent the limiting condition used by Huang et al. The substrate conductivity(k$_s$) is used as a normalizing factor for both of them. The results of this simulation is given in figure 3. 

\begin {figure}[h!]
 \includegraphics[width=1\linewidth]{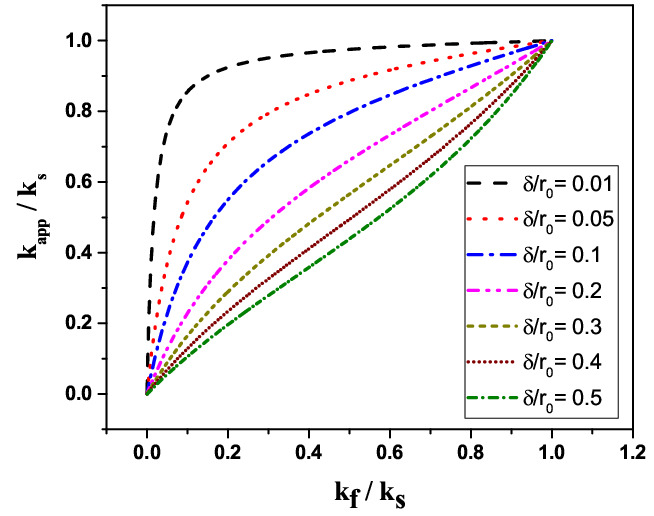}
\caption{ Relationship pattern between $k_{app}$ and $k_f$ for various values of $\frac{\delta}{r_0}$.}
\end{figure}

It is observed from figure 3 that for $\delta/r_o$ ratio of 0.01 (very thin film region), curve shows a highly concave  nature and the effect of $k_{app}$ is predominant over $k_f$. The concavity of the curve decreases for thicker films as $\delta/r_o$ value increases to 0.4. A reversal of nature is observed at $\delta/r_o$ =0.4. This marks transition of film from thin film to bulk and can be considered as the critical value for the associated material beyond which assumptions pertaining to thin film limit will not work. This can be explained by the fact that $\delta/r_o$ ratio increases with increase in film thickness. As a result, heat flow across the film increases thereby increasing significance of  $k_f$ parameter. In very thin films, heat flow occurs beyond the film in entire sample making $k_{app}$ value significant. Conversely, for bulk films, heat transfer is limited mostly to the film making $k_f$ value significant. Simulation of equation 5 provides us a generous plot to observe this transition as we move from thin to bulk region as presented in figure 3. \\

\subsection{Calculation of thermal conductivity of 10H SiC films}
 Thermal conductivity of 10H  SiC films were calculated by  utilising equation 4 and 5 of section IV A. It is known that thermal conductivity is a function of temperature of the film when it is subjected to a heat gradient. In the current study, shift in Raman peak position is indicative of thermal gradient of the film. The thermal gradient was achieved by irradiating the film with a 532 nm laser at two different laser power i.e 1.25 mW and 5 mW.\ 
 Raman spectra of films were first measured by keeping laser power at 1.25 mW, wherein laser induced sample heating is presumed to be negligible. Sample heating was then carried out by increasing the temperature from 298K to 773K using a Linkam stage and measuring the corresponding Raman spectrum. Raman spectra thus obtained had a Raman peak at 960cm$^{-1}$, which is a characteristic of hexagonal SiC polytype.\cite{Vlasta} The shift in this 960 cm$^{-1}$ peak for 156 nm 10H SiC film with respect to change in temperature is shown in fig 4. It is observed that with increase in temperature, peak shifted towards lower wave number(red shift). The red shift is due to anharmonicity induced in the film with increase in temperature. Similar red shift was also observed for other two films. Subsequently, a calibration plot was generated by plotting Raman peak shift against temperature and is shown in fig 5. From the linear plot one can observe that with increase in substrate temperature the Raman peak shift reduced. The slope became shallower with higher thickness, indicating a smaller temperature difference between two  consecutive points. Subsequently, with higher thickness the value of $\Delta$T reduces and value of k$_{app}$ increases. A separate set of Raman spectra were recorded at room temperature by keeping laser power at 5mW for deducing apparent thermal conductivity (k$_{app}$). The sample heating due to change in laser power was determined by observing the change in peak position and subsequently utilising the calibration plot to find the corresponding temperature change in the film, which is $\Delta$T. This temperature difference is then used in equation 4 of section IV A to calculate the apparent thermal conductivity of the sample (k$_{app}$), which includes thermal conductivity of film and substrate.  Since, thickness($\delta$) for each film is known from cross-sectional FESEM images, we calculated thermal conductivity of the film (K$_f$) by using equation 5 of section IV A. Two roots were obtained for the quadratic equation in each case. The negative root was ignored on account of physical impossibility and only the positive root for K$_f$ is reported in Table 1.  \\

\begin{figure}[htbp]
\begin{center}
\includegraphics[width=1\linewidth]{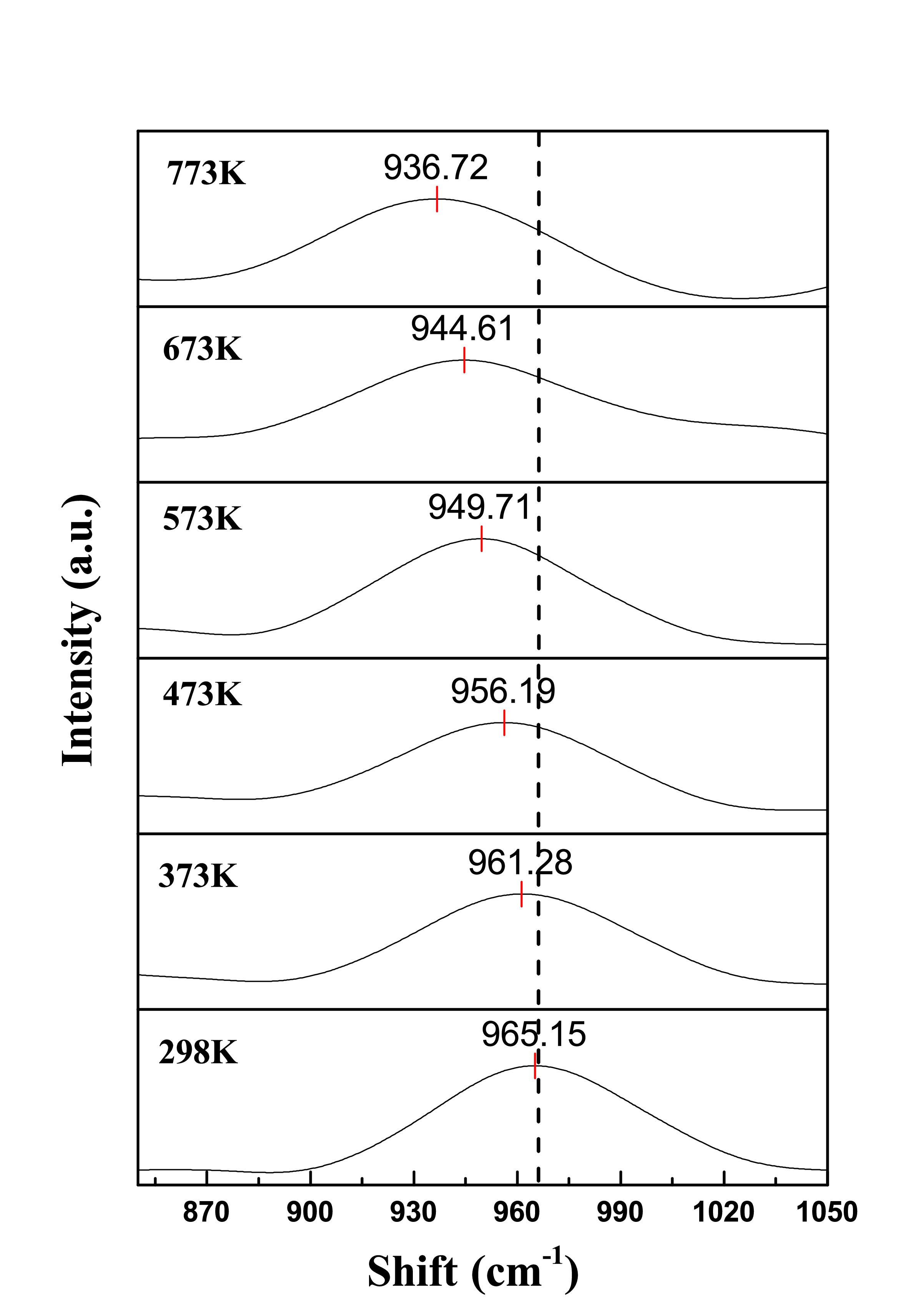}
 \end{center}
\caption{ SiC Raman band shift with temperature of 156 nm thick 10H-SiC film measured irradiated with 1.25 mW laser power}
\end{figure}
As presented in Table 1 $\delta/r_o$ values for films of thickness 104, 135 and 156 nm are 0.18, 0.23 and 0.26, respectively. These values lie in thin film limit as indicated in fig 3. It can be observed in fig 3 that when $\delta/r_o$ varies from 0.2 to 0.3, the relationship between K$_{app}$ and K$_f$ tends to become linear implying  K$_{app}$ becomes equal to K$_f$. This is corroborated in our calculation. It can be observed in table 1 that with increase in $\delta/r_o$ value from 0.18 to 0.26, K$_{app}$ becomes comparable to K$_f$ for 156 nm film.

\begin{figure}[ht] 
\includegraphics[width=1\linewidth]{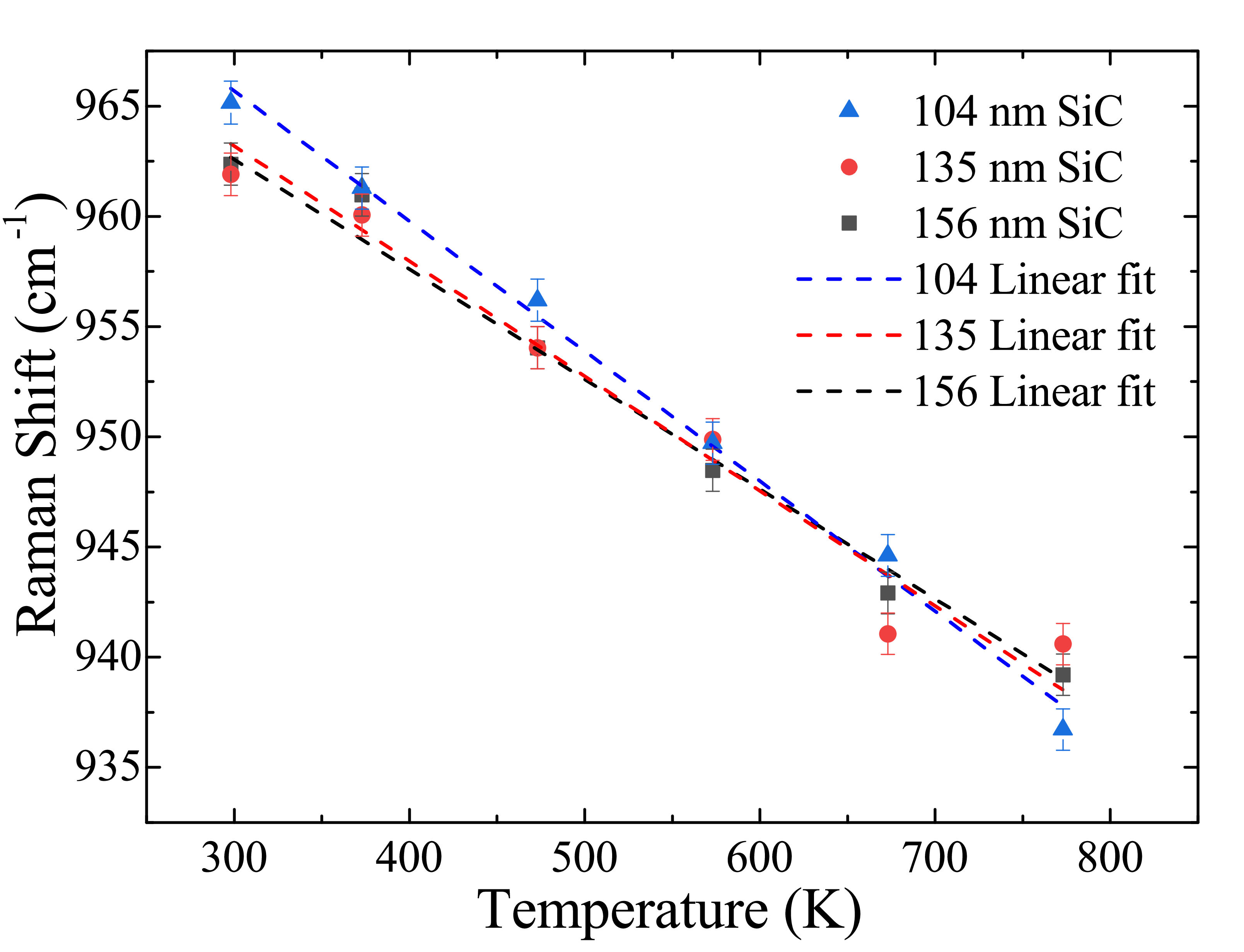}
\caption{Raman Shift Vs Temperature calibration plots of 104, 135, and 156 nm 10H-SiC films}
\end{figure}

\begin{table}[htbp]
\caption{${\delta}$/$r_o$, calculated K$_{app}$ and K$_f$ values for 10H SiC films of different thicknesses}
\begin{tabular}{|c|c|c|c|c|}
\hline
\begin{tabular}[c]{@{}c@{}}Thickness\\ (nm)\end{tabular} & $\delta/r_0$   & $\Delta$ T & \begin{tabular}[c]{@{}c@{}}$K_{app}$\\ (using eq. 4)\end{tabular} & \begin{tabular}[c]{@{}c@{}}$K_f$\\ $(Wm^{-1}K^{-1})$\\ {[}using eq. 5{]}\end{tabular} \\ \hline
104                                                      & 0.18  & 36.79  & 33.81                                                       & 11.87                                                                     \\ \hline
135                                                      & 0.23   & 32.69  & 38.05                                                       & 17.86                                                                     \\ \hline
156                                                      & 0.26 & 10.97  & 113.32                                                      & 102.36                                                                    \\ \hline
\end{tabular}
\end{table}
This also indicates that within thin film limit, thermal conductivity value can vary within range of 11- 103 $(Wm^{-1}K^{-1})$. This gives us an idea about the range of thermal conductivity values for 10H SiC films below 200 nm thickness. There have been various reports on thermal  conductivity of SiC. Slack et al. observed that thermal conductivity of SiC monocrystal at room temperature is 490 $(Wm^{-1}K^{-1})$. \cite{slack1964thermal} 

Composite  materials in SiC  was found to have reduced thermal conductivity value of 252–270 $(Wm^{-1}K^{-1})$ as compared to monocrystals studied by slack et al.\cite{Nakano2004} However there is no report of thermal conductivity of phase pure SiC thin film in any  polytype. So, there is no reference to compare thermal conductivity of thin film (K$_f$) and we are hereby reporting and effective way to determine thermal conductivity of phase pure SiC thin films by a nondestructive method  for the first time. As per our observation, thermal conductivity of 10H SiC is 102.36 $(Wm^{-1}K^{-1})$.   \\

\section{CONCLUSION}

 In this work, we have calculated thermal conductivity of 10H SiC films of three thicknesses by utilising Raman Thermometry. A novel thin film limit was worked out for films in nanometer range by extrapolating the model given by Huang et al. and the same was applied to calculate thermal conductivity of 10H SiC film. Our observation confirms thickness dependence of thermal conductivity which can vary from 11 -102 $(Wm^{-1}K^{-1})$.  It was found that the 156 nm film is the ideal thickness for 10H SiC films to calculate thermal conductivity by this method as its K$_{app}$ is comparable to K$_f$.  \\

\begin{acknowledgments}
Madhusmita Sahoo acknowledges Dr Shaju Albert for his support in carrying out the project. The authors from NISER acknowledge the Department of Atomic Energy (DAE), India for supporting this work through the project RIN-4001.\\

\section*{Statements and Declarations}
The authors declare that they have no known competing financial interests or personal relationships that could have appeared to influence the work reported in this paper.

\end{acknowledgments}
\vspace{-0.3in}

\nocite{*}
\bibliography{SiC.bib}

\begin{thebibliography}{18}%
\makeatletter
\providecommand \@ifxundefined [1]{%
 \@ifx{#1\undefined}
}%
\providecommand \@ifnum [1]{%
 \ifnum #1\expandafter \@firstoftwo
 \else \expandafter \@secondoftwo
 \fi
}%
\providecommand \@ifx [1]{%
 \ifx #1\expandafter \@firstoftwo
 \else \expandafter \@secondoftwo
 \fi
}%
\providecommand \natexlab [1]{#1}%
\providecommand \enquote  [1]{``#1''}%
\providecommand \bibnamefont  [1]{#1}%
\providecommand \bibfnamefont [1]{#1}%
\providecommand \citenamefont [1]{#1}%
\providecommand \href@noop [0]{\@secondoftwo}%
\providecommand \href [0]{\begingroup \@sanitize@url \@href}%
\providecommand \@href[1]{\@@startlink{#1}\@@href}%
\providecommand \@@href[1]{\endgroup#1\@@endlink}%
\providecommand \@sanitize@url [0]{\catcode `\\12\catcode `\$12\catcode
  `\&12\catcode `\#12\catcode `\^12\catcode `\_12\catcode `\%12\relax}%
\providecommand \@@startlink[1]{}%
\providecommand \@@endlink[0]{}%
\providecommand \url  [0]{\begingroup\@sanitize@url \@url }%
\providecommand \@url [1]{\endgroup\@href {#1}{\urlprefix }}%
\providecommand \urlprefix  [0]{URL }%
\providecommand \Eprint [0]{\href }%
\providecommand \doibase [0]{http://dx.doi.org/}%
\providecommand \selectlanguage [0]{\@gobble}%
\providecommand \bibinfo  [0]{\@secondoftwo}%
\providecommand \bibfield  [0]{\@secondoftwo}%
\providecommand \translation [1]{[#1]}%
\providecommand \BibitemOpen [0]{}%
\providecommand \bibitemStop [0]{}%
\providecommand \bibitemNoStop [0]{.\EOS\space}%
\providecommand \EOS [0]{\spacefactor3000\relax}%
\providecommand \BibitemShut  [1]{\csname bibitem#1\endcsname}%
\let\auto@bib@innerbib\@empty
\bibitem [{\citenamefont {Nava}\ \emph {et~al.}(2008)\citenamefont {Nava},
  \citenamefont {Bertuccio}, \citenamefont {Cavallini},\ and\ \citenamefont
  {Vittone}}]{nava2008silicon}%
  \BibitemOpen
  \bibfield  {author} {\bibinfo {author} {\bibfnamefont {F.}~\bibnamefont
  {Nava}}, \bibinfo {author} {\bibfnamefont {G.}~\bibnamefont {Bertuccio}},
  \bibinfo {author} {\bibfnamefont {A.}~\bibnamefont {Cavallini}}, \ and\
  \bibinfo {author} {\bibfnamefont {E.}~\bibnamefont {Vittone}},\ }\bibfield
  {title} {\enquote {\bibinfo {title} {Silicon carbide and its use as a
  radiation detector material},}\ }\href@noop {} {\bibfield  {journal}
  {\bibinfo  {journal} {Measurement Science and Technology}\ }\textbf {\bibinfo
  {volume} {19}},\ \bibinfo {pages} {102001} (\bibinfo {year}
  {2008})}\BibitemShut {NoStop}%
\bibitem [{\citenamefont {Cheng~Z.}(2022)}]{Cheng2022}%
  \BibitemOpen
  \bibfield  {author} {\bibinfo {author} {\bibfnamefont {K.~K. e.~a.}\
  \bibnamefont {Cheng~Z.}, \bibfnamefont {Liang~J.}},\ }\bibfield  {title}
  {\enquote {\bibinfo {title} {High thermal conductivity in wafer-scale cubic
  silicon carbide crystals.}}\ }\href@noop {} {\bibfield  {journal} {\bibinfo
  {journal} {Nature Communications}\ }\textbf {\bibinfo {volume} {13}},\
  \bibinfo {pages} {7201} (\bibinfo {year} {2022})}\BibitemShut {NoStop}%
\bibitem [{\citenamefont {Slack}(1964)}]{slack1964thermal}%
  \BibitemOpen
  \bibfield  {author} {\bibinfo {author} {\bibfnamefont {G.~A.}\ \bibnamefont
  {Slack}},\ }\bibfield  {title} {\enquote {\bibinfo {title} {Thermal
  conductivity of pure and impure silicon, silicon carbide, and diamond},}\
  }\href@noop {} {\bibfield  {journal} {\bibinfo  {journal} {Journal of Applied
  physics}\ }\textbf {\bibinfo {volume} {35}},\ \bibinfo {pages} {3460--3466}
  (\bibinfo {year} {1964})}\BibitemShut {NoStop}%
\bibitem [{\citenamefont {Protik}\ \emph {et~al.}(2017)\citenamefont {Protik},
  \citenamefont {Katre}, \citenamefont {Lindsay}, \citenamefont {Carrete},
  \citenamefont {Mingo},\ and\ \citenamefont {Broido}}]{PROTIK201731}%
  \BibitemOpen
  \bibfield  {author} {\bibinfo {author} {\bibfnamefont {N.~H.}\ \bibnamefont
  {Protik}}, \bibinfo {author} {\bibfnamefont {A.}~\bibnamefont {Katre}},
  \bibinfo {author} {\bibfnamefont {L.}~\bibnamefont {Lindsay}}, \bibinfo
  {author} {\bibfnamefont {J.}~\bibnamefont {Carrete}}, \bibinfo {author}
  {\bibfnamefont {N.}~\bibnamefont {Mingo}}, \ and\ \bibinfo {author}
  {\bibfnamefont {D.}~\bibnamefont {Broido}},\ }\bibfield  {title} {\enquote
  {\bibinfo {title} {Phonon thermal transport in 2h, 4h and 6h silicon carbide
  from first principles},}\ }\href {\doibase
  https://doi.org/10.1016/j.mtphys.2017.05.004} {\bibfield  {journal} {\bibinfo
   {journal} {Materials Today Physics}\ }\textbf {\bibinfo {volume} {1}},\
  \bibinfo {pages} {31--38} (\bibinfo {year} {2017})}\BibitemShut {NoStop}%
\bibitem [{\citenamefont {Zhang}\ and\ \citenamefont
  {Grigoropoulos}(1995)}]{zhang1995thermal}%
  \BibitemOpen
  \bibfield  {author} {\bibinfo {author} {\bibfnamefont {X.}~\bibnamefont
  {Zhang}}\ and\ \bibinfo {author} {\bibfnamefont {C.~P.}\ \bibnamefont
  {Grigoropoulos}},\ }\bibfield  {title} {\enquote {\bibinfo {title} {Thermal
  conductivity and diffusivity of free-standing silicon nitride thin films},}\
  }\href@noop {} {\bibfield  {journal} {\bibinfo  {journal} {Review of
  scientific instruments}\ }\textbf {\bibinfo {volume} {66}},\ \bibinfo {pages}
  {1115--1120} (\bibinfo {year} {1995})}\BibitemShut {NoStop}%
\bibitem [{\citenamefont {Cahill}, \citenamefont {Katiyar},\ and\ \citenamefont
  {Abelson}(1994)}]{cahill1994thermal}%
  \BibitemOpen
  \bibfield  {author} {\bibinfo {author} {\bibfnamefont {D.~G.}\ \bibnamefont
  {Cahill}}, \bibinfo {author} {\bibfnamefont {M.}~\bibnamefont {Katiyar}}, \
  and\ \bibinfo {author} {\bibfnamefont {J.}~\bibnamefont {Abelson}},\
  }\bibfield  {title} {\enquote {\bibinfo {title} {Thermal conductivity of
  a-si: H thin films},}\ }\href@noop {} {\bibfield  {journal} {\bibinfo
  {journal} {Physical review B}\ }\textbf {\bibinfo {volume} {50}},\ \bibinfo
  {pages} {6077} (\bibinfo {year} {1994})}\BibitemShut {NoStop}%
\bibitem [{\citenamefont {Swimm}(1983)}]{swimm1983photoacoustic}%
  \BibitemOpen
  \bibfield  {author} {\bibinfo {author} {\bibfnamefont {R.~T.}\ \bibnamefont
  {Swimm}},\ }\bibfield  {title} {\enquote {\bibinfo {title} {Photoacoustic
  determination of thin-film thermal properties},}\ }\href@noop {} {\bibfield
  {journal} {\bibinfo  {journal} {Applied Physics Letters}\ }\textbf {\bibinfo
  {volume} {42}},\ \bibinfo {pages} {955--957} (\bibinfo {year}
  {1983})}\BibitemShut {NoStop}%
\bibitem [{\citenamefont {Callard}\ \emph {et~al.}(1999)\citenamefont
  {Callard}, \citenamefont {Tallarida}, \citenamefont {Borghesi},\ and\
  \citenamefont {Zanotti}}]{callard1999thermal}%
  \BibitemOpen
  \bibfield  {author} {\bibinfo {author} {\bibfnamefont {S.}~\bibnamefont
  {Callard}}, \bibinfo {author} {\bibfnamefont {G.}~\bibnamefont {Tallarida}},
  \bibinfo {author} {\bibfnamefont {A.}~\bibnamefont {Borghesi}}, \ and\
  \bibinfo {author} {\bibfnamefont {L.}~\bibnamefont {Zanotti}},\ }\bibfield
  {title} {\enquote {\bibinfo {title} {Thermal conductivity of sio2 films by
  scanning thermal microscopy},}\ }\href@noop {} {\bibfield  {journal}
  {\bibinfo  {journal} {Journal of non-crystalline solids}\ }\textbf {\bibinfo
  {volume} {245}},\ \bibinfo {pages} {203--209} (\bibinfo {year}
  {1999})}\BibitemShut {NoStop}%
\bibitem [{\citenamefont {Burzo}, \citenamefont {Komarov},\ and\ \citenamefont
  {Raad}(2002)}]{burzo2002influence}%
  \BibitemOpen
  \bibfield  {author} {\bibinfo {author} {\bibfnamefont {M.~G.}\ \bibnamefont
  {Burzo}}, \bibinfo {author} {\bibfnamefont {P.~L.}\ \bibnamefont {Komarov}},
  \ and\ \bibinfo {author} {\bibfnamefont {P.~E.}\ \bibnamefont {Raad}},\
  }\bibfield  {title} {\enquote {\bibinfo {title} {Influence of the metallic
  absorption layer on the quality of thermal conductivity measurements by the
  transient thermo-reflectance method},}\ }\href@noop {} {\bibfield  {journal}
  {\bibinfo  {journal} {Microelectronics journal}\ }\textbf {\bibinfo {volume}
  {33}},\ \bibinfo {pages} {697--703} (\bibinfo {year} {2002})}\BibitemShut
  {NoStop}%
\bibitem [{\citenamefont {Komarov}\ \emph {et~al.}(2003)\citenamefont
  {Komarov}, \citenamefont {Burzo}, \citenamefont {Kaytaz},\ and\ \citenamefont
  {Raad}}]{komarov2003transient}%
  \BibitemOpen
  \bibfield  {author} {\bibinfo {author} {\bibfnamefont {P.~L.}\ \bibnamefont
  {Komarov}}, \bibinfo {author} {\bibfnamefont {M.~G.}\ \bibnamefont {Burzo}},
  \bibinfo {author} {\bibfnamefont {G.}~\bibnamefont {Kaytaz}}, \ and\ \bibinfo
  {author} {\bibfnamefont {P.~E.}\ \bibnamefont {Raad}},\ }\bibfield  {title}
  {\enquote {\bibinfo {title} {Transient thermo-reflectance measurements of the
  thermal conductivity and interface resistance of metallized natural and
  isotopically-pure silicon},}\ }\href@noop {} {\bibfield  {journal} {\bibinfo
  {journal} {Microelectronics journal}\ }\textbf {\bibinfo {volume} {34}},\
  \bibinfo {pages} {1115--1118} (\bibinfo {year} {2003})}\BibitemShut {NoStop}%
\bibitem [{\citenamefont {P{\'e}richon}\ \emph {et~al.}(2000)\citenamefont
  {P{\'e}richon}, \citenamefont {Lysenko}, \citenamefont {Roussel},
  \citenamefont {Remaki}, \citenamefont {Champagnon}, \citenamefont {Barbier},\
  and\ \citenamefont {Pinard}}]{perichon2000}%
  \BibitemOpen
  \bibfield  {author} {\bibinfo {author} {\bibfnamefont {S.}~\bibnamefont
  {P{\'e}richon}}, \bibinfo {author} {\bibfnamefont {V.}~\bibnamefont
  {Lysenko}}, \bibinfo {author} {\bibfnamefont {P.}~\bibnamefont {Roussel}},
  \bibinfo {author} {\bibfnamefont {B.}~\bibnamefont {Remaki}}, \bibinfo
  {author} {\bibfnamefont {B.}~\bibnamefont {Champagnon}}, \bibinfo {author}
  {\bibfnamefont {D.}~\bibnamefont {Barbier}}, \ and\ \bibinfo {author}
  {\bibfnamefont {P.}~\bibnamefont {Pinard}},\ }\bibfield  {title} {\enquote
  {\bibinfo {title} {Technology and micro-raman characterization of thick
  meso-porous silicon layers for thermal effect microsystems},}\ }\href@noop {}
  {\bibfield  {journal} {\bibinfo  {journal} {Sensors and Actuators A:
  Physical}\ }\textbf {\bibinfo {volume} {85}},\ \bibinfo {pages} {335--339}
  (\bibinfo {year} {2000})}\BibitemShut {NoStop}%
\bibitem [{\citenamefont {Huang~S.}(2009)}]{Huang2009}%
  \BibitemOpen
  \bibfield  {author} {\bibinfo {author} {\bibfnamefont {F.~X. e.~a.}\
  \bibnamefont {Huang~S.}, \bibfnamefont {Ruan~Xd.}},\ }\bibfield  {title}
  {\enquote {\bibinfo {title} {Measurement of the thermal transport properties
  of dielectric thin films using the micro-raman method.}}\ }\href@noop {}
  {\bibfield  {journal} {\bibinfo  {journal} {J. Zhejiang Univ. Sci. A}\
  }\textbf {\bibinfo {volume} {10}} (\bibinfo {year} {2009})}\BibitemShut
  {NoStop}%
\bibitem [{\citenamefont {Gan}, \citenamefont {Samvedi},\ and\ \citenamefont
  {Tomar}(2015)}]{gan2015raman}%
  \BibitemOpen
  \bibfield  {author} {\bibinfo {author} {\bibfnamefont {M.}~\bibnamefont
  {Gan}}, \bibinfo {author} {\bibfnamefont {V.}~\bibnamefont {Samvedi}}, \ and\
  \bibinfo {author} {\bibfnamefont {V.}~\bibnamefont {Tomar}},\ }\bibfield
  {title} {\enquote {\bibinfo {title} {Raman spectroscopy-based investigation
  of thermal conductivity of stressed silicon microcantilevers},}\ }\href@noop
  {} {\bibfield  {journal} {\bibinfo  {journal} {Journal of Thermophysics and
  Heat Transfer}\ }\textbf {\bibinfo {volume} {29}},\ \bibinfo {pages}
  {845--857} (\bibinfo {year} {2015})}\BibitemShut {NoStop}%
\bibitem [{\citenamefont {Freedman}\ \emph {et~al.}(2017)\citenamefont
  {Freedman}, \citenamefont {Goyal}, \citenamefont {Ahn},\ and\ \citenamefont
  {Kim}}]{freedman2017substrate}%
  \BibitemOpen
  \bibfield  {author} {\bibinfo {author} {\bibfnamefont {K.~J.}\ \bibnamefont
  {Freedman}}, \bibinfo {author} {\bibfnamefont {G.}~\bibnamefont {Goyal}},
  \bibinfo {author} {\bibfnamefont {C.~W.}\ \bibnamefont {Ahn}}, \ and\
  \bibinfo {author} {\bibfnamefont {M.~J.}\ \bibnamefont {Kim}},\ }\bibfield
  {title} {\enquote {\bibinfo {title} {Substrate dependent ad-atom migration on
  graphene and the impact on electron-beam sculpting functional nanopores},}\
  }\href@noop {} {\bibfield  {journal} {\bibinfo  {journal} {Sensors}\ }\textbf
  {\bibinfo {volume} {17}},\ \bibinfo {pages} {1091} (\bibinfo {year}
  {2017})}\BibitemShut {NoStop}%
\bibitem [{\citenamefont {Zhang}, \citenamefont {Gan},\ and\ \citenamefont
  {Tomar}(2014)}]{zhang2014raman}%
  \BibitemOpen
  \bibfield  {author} {\bibinfo {author} {\bibfnamefont {Y.}~\bibnamefont
  {Zhang}}, \bibinfo {author} {\bibfnamefont {M.}~\bibnamefont {Gan}}, \ and\
  \bibinfo {author} {\bibfnamefont {V.}~\bibnamefont {Tomar}},\ }\bibfield
  {title} {\enquote {\bibinfo {title} {Raman thermometry based thermal
  conductivity measurement of bovine cortical bone as a function of compressive
  stress},}\ }\href@noop {} {\bibfield  {journal} {\bibinfo  {journal} {Journal
  of Nanotechnology in Engineering and Medicine}\ }\textbf {\bibinfo {volume}
  {5}} (\bibinfo {year} {2014})}\BibitemShut {NoStop}%
\bibitem [{\citenamefont {Dryden}(1983)}]{Dryden}%
  \BibitemOpen
  \bibfield  {author} {\bibinfo {author} {\bibfnamefont {J.~R.}\ \bibnamefont
  {Dryden}},\ }\bibfield  {title} {\enquote {\bibinfo {title} {{The Effect of a
  Surface Coating on the Constriction Resistance of a Spot on an Infinite
  Half-Plane}},}\ }\href {\doibase 10.1115/1.3245596} {\bibfield  {journal}
  {\bibinfo  {journal} {Journal of Heat Transfer}\ }\textbf {\bibinfo {volume}
  {105}},\ \bibinfo {pages} {408--410} (\bibinfo {year} {1983})},\ \Eprint
  {http://arxiv.org/abs/https://asmedigitalcollection.asme.org/heattransfer/article-pdf/105/2/408/5573361/408\_1.pdf}
  {https://asmedigitalcollection.asme.org/heattransfer/article-pdf/105/2/408/5573361/408\_1.pdf}
  \BibitemShut {NoStop}%
\bibitem [{\citenamefont {Sasinkov{\'a}}\ \emph {et~al.}(2015)\citenamefont
  {Sasinkov{\'a}}, \citenamefont {Huran}, \citenamefont {Kleinov{\'a}},
  \citenamefont {Boh{\'a}ček}, \citenamefont {Arbet},\ and\ \citenamefont
  {Sek{\'a}čova}}]{Vlasta}%
  \BibitemOpen
  \bibfield  {author} {\bibinfo {author} {\bibfnamefont {V.}~\bibnamefont
  {Sasinkov{\'a}}}, \bibinfo {author} {\bibfnamefont {J.}~\bibnamefont
  {Huran}}, \bibinfo {author} {\bibfnamefont {A.}~\bibnamefont {Kleinov{\'a}}},
  \bibinfo {author} {\bibfnamefont {P.}~\bibnamefont {Boh{\'a}ček}}, \bibinfo
  {author} {\bibfnamefont {J.}~\bibnamefont {Arbet}}, \ and\ \bibinfo {author}
  {\bibfnamefont {M.}~\bibnamefont {Sek{\'a}čova}},\ }\bibfield  {title}
  {\enquote {\bibinfo {title} {{Raman spectroscopy study of SiC thin films
  prepared by PECVD for solar cell working in hard environment}},}\ }in\ \href
  {\doibase 10.1117/12.2186749} {\emph {\bibinfo {booktitle} {Reliability of
  Photovoltaic Cells, Modules, Components, and Systems VIII}}},\ Vol.\ \bibinfo
  {volume} {9563},\ \bibinfo {editor} {edited by\ \bibinfo {editor}
  {\bibfnamefont {N.~G.}\ \bibnamefont {Dhere}}, \bibinfo {editor}
  {\bibfnamefont {J.~H.}\ \bibnamefont {Wohlgemuth}}, \ and\ \bibinfo {editor}
  {\bibfnamefont {R.}~\bibnamefont {Jones-Albertus}}},\ \bibinfo {organization}
  {International Society for Optics and Photonics}\ (\bibinfo  {publisher}
  {SPIE},\ \bibinfo {year} {2015})\ p.\ \bibinfo {pages} {95630V}\BibitemShut
  {NoStop}%
\bibitem [{\citenamefont {Nakano}\ \emph {et~al.}(2004)\citenamefont {Nakano},
  \citenamefont {Watari}, \citenamefont {Kinemuchi}, \citenamefont {Ishizaki},\
  and\ \citenamefont {Urabe}}]{Nakano2004}%
  \BibitemOpen
  \bibfield  {author} {\bibinfo {author} {\bibfnamefont {H.}~\bibnamefont
  {Nakano}}, \bibinfo {author} {\bibfnamefont {K.}~\bibnamefont {Watari}},
  \bibinfo {author} {\bibfnamefont {Y.}~\bibnamefont {Kinemuchi}}, \bibinfo
  {author} {\bibfnamefont {K.}~\bibnamefont {Ishizaki}}, \ and\ \bibinfo
  {author} {\bibfnamefont {K.}~\bibnamefont {Urabe}},\ }\bibfield  {title}
  {\enquote {\bibinfo {title} {Microstructural characterization of
  high-thermal-conductivity {SiC} ceramics},}\ }\href {\doibase
  10.1016/j.jeurceramsoc.2003.12.019} {\bibfield  {journal} {\bibinfo
  {journal} {Journal of the European Ceramic Society}\ }\textbf {\bibinfo
  {volume} {24}},\ \bibinfo {pages} {3685--3690} (\bibinfo {year}
  {2004})}\BibitemShut {NoStop}%
\end{thebibliography}%

\end{document}